\documentclass[%
 reprint,
superscriptaddress,
preprintnumbers,
nofootinbib,
 amsmath,amssymb,
 aps,
]{revtex4-1}

\pdfoutput=1

\usepackage[colorlinks]{hyperref}
\usepackage{graphicx}
\usepackage{dcolumn}
\usepackage{bm}
\usepackage{color}
\usepackage{slashed}
\usepackage[dvipsnames]{xcolor}
\usepackage{placeins}
\usepackage{slashed}
\usepackage{multirow}
\usepackage{subfigure}
\usepackage{tabularx}
\usepackage{mathtools}
\usepackage{floatrow}
\usepackage{blindtext}
\usepackage{soul}
\usepackage{slashed}
\usepackage{amsmath}
\usepackage{braket}
\usepackage{comment}

\hypersetup{
  colorlinks=true,
  linkcolor=red,
  citecolor=blue,
  urlcolor=blue,
}

\definecolor{coral}{RGB}{255,127,80}
\definecolor{indigo}{RGB}{75,0,130}
\definecolor{red}{rgb}{0.9, 0,0}
\definecolor{cerulean}{rgb}{0., 0.62,0.9}
\definecolor{navy}{rgb}{0.05, 0.05,0.8}

\newfloatcommand{capbtabbox}{table}[][\FBwidth]

\newcommand{\GeV}{{\, \rm GeV}}
\newcommand{\MeV}{{\, \rm MeV}}

\newcommand{\be}{\begin{equation}}
\newcommand{\ee}{\end{equation}}

\begin{document}

\preprint{TTP23-045, P3H-23-071, MPP-2023-240, FR-PHENO-2023-11}
\title{A Leptonic ALP Portal to the Dark Sector}

\author{Giovanni Armando}
\affiliation{Dipartimento di Fisica E. Fermi, Universit\`a di Pisa, Largo B. Pontecorvo 3, I-56127 Pisa, Italy }
\affiliation{INFN, Sezione di Pisa, Largo Bruno Pontecorvo 3, I-56127 Pisa, Italy}
\author{Paolo Panci}
\affiliation{Dipartimento di Fisica E. Fermi, Universit\`a di Pisa, Largo B. Pontecorvo 3, I-56127 Pisa, Italy }
\affiliation{INFN, Sezione di Pisa, Largo Bruno Pontecorvo 3, I-56127 Pisa, Italy}
\author{Joachim Weiss}
\affiliation{Max Planck Institute for Physics,
F\"ohringer Ring 6, 80805 M\"unchen, Germany}
\affiliation{Technische Universit\"at M\"unchen, Physik-Department, James-Franck-Strasse 1, 85748 Garching, Germany}
\author{Robert Ziegler}
\affiliation{Institut f\"ur Theoretische Teilchenphysik, Karlsruhe Institute of Technology, Karlsruhe, Germany}
\affiliation{Physikalisches Institut, Albert-Ludwigs-Universit\"at Freiburg,
79104 Freiburg, Germany}
\date{\today}

\begin{abstract}

We discuss the leptonic axion-like particle (ALP) portal as a simple scenario that connects observed discrepancies in anomalous magnetic moments to the dark matter relic abundance. In this framework an axion-like particle in the multi-MeV range couples to SM leptons and a dark matter (DM) fermion, with mass above the ALP mass but below 1 GeV. The ALP contributes to $(g-2)_\mu$ and  $(g-2)_e$ dominantly through two-loop Barr-Zee diagrams, while the DM abundance is generated by p-wave annihilation to ALP pairs. Constraints from beam-dump experiments, colliders, and cosmic microwave background probes are very stringent, and restrict the viable parameter space to a rather narrow region that will be tested in the near future.

\end{abstract}

\maketitle

\section{Introduction}\label{sec:intro}

The origin of Dark Matter (DM) is arguably the most pressing problem of contemporary particle physics. Within generic theories beyond the Standard Model (BSM), the best motivated DM   candidates are those that naturally arise in scenarios addressing  other problems and shortcomings of the Standard Model (SM), for example the Strong CP Problem (as the QCD axion), or the hierarchy problem (as the neutralino). Another interesting class of models aims to generate the DM relic abundance within BSM scenarios that address experimental anomalies, i.e. explain observed deviations from SM predictions. Here we focus on the longstanding discrepancy in anomalous magnetic moments of leptons $(g-2)_\ell$, which have been addressed in a variety of BSM scenarios, see \cite{Athron:2021iuf} for a review. 

The possible connection of DM and $(g-2)_\ell$ has often been considered in the context of heavy new particles with masses ${\cal O}(100 \GeV)$,  see e.g. Ref.~\cite{Calibbi:2018rzv, De:2021crr}. In this article instead we analyze the scenario where only light new particles in the multi-MeV range are present: a pseudo-scalar particle coupling only to leptons and a new SM singlet fermion that accounts for DM. The lightness of the fermion is ensured by the chiral symmetry, while the scalar is light because it arises as a pseudo-Goldstone boson of an approximate Peccei-Quinn (PQ) symmetry, usually referred  to as an axion-like particle (ALP). This ALP acts as a mediator between the SM leptons  and the dark fermion, and gives rise to the observed DM relic abundance via thermal freeze-out.   

Similar scenarios have been discussed in previous works~\cite{Nomura:2008ru, Freytsis:2010ne, Dolan:2014ska, Berlin:2015wwa, Fan:2015sza, No:2015xqa, Bauer:2017ota, Baek:2017vzd, Kamada:2017tsq, Kaneta:2017wfh, Banerjee:2017wxi, Arcadi:2017wqi, Hochberg:2018rjs, Berlin:2018bsc, deNiverville:2019xsx, Darme:2020sjf, Ge:2021cjz, Gola:2021abm, Domcke:2021yuz, Zhevlakov:2022vio, Bauer:2022rwf, Bharucha:2022lty,  Fitzpatrick:2023xks, Ghosh:2023tyz,Dror:2023fyd, Capozzi:2023ffu}, often in the  context of Higgs-ALP mixing, i.e., the ALP that inherits all Higgs couplings to fermions, suppressed by a mixing angle. Here instead we only consider ALP couplings to leptons, all taken as independent parameters, and focus on a relatively low value of the ALP decay constant $f_a$. Our region of parameter space actually resembles the ``visible QCD" axion proposed in Ref.~\cite{Alves:2017avw,Alves:2020xhf}, which is a QCD axion with decay constant in the GeV range and couplings only to first-generation fermions. Here  instead the ALP couples to all three leptons, while couplings to quarks are completely absent. This renders this model phenomenologically viable, in contrast to the visible QCD axion of Ref.~\cite{Alves:2017avw}, which  is largely (if not completely\footnote{The theoretical prediction might be subject to additional suppression factors, depending on different assumptions on the leading ChPT operator, see Ref.~\cite{Hostert:2020xku} }) excluded by NA62 searches~\cite{NA62:2023rvm} for $K \to \pi a a$, with all ALPs promptly decaying to electrons. 

A very similar model to the one considered here has been employed in Ref.~\cite{Buttazzo:2020vfs} to simultaneously explain $(g-2)_\ell$ and an excess of electron events observed in the XENON1T experiment~\cite{XENON:2020rca}, induced by the scattering of an asymmetric DM fermion. This anomaly has now been refuted by additional data~\cite{XENON:2022ltv}, along with new bounds recently derived for ALP-electron couplings from Kaon decays~\cite{Altmannshofer:2022ckw}. Here we take a somewhat smaller ALP-electron coupling in order to satisfy the latter constraints, and consider thermal freeze-out instead of an asymmetric DM scenario to generate the DM relic abundance.

The particle phenomenology in our scenario is controlled by the ALP and its couplings to leptons, since ALP decays to the dark sector are kinematically closed. As $m_a < m_\mu$,  only ALP couplings to electrons and photons are relevant, the latter induced by loops of SM fermions. This allows to populate also regions of parameter space at lower ALP masses ($m_a \lesssim 30 \MeV$) deemed to be excluded in Ref.~\cite{Buen-Abad:2021fwq}, which analyzed generic light ALP explanations of $(g-2)_\mu$. Nevertheless we share the conclusion of these authors that it is challenging to build UV-complete models of these scenarios, because the PQ breaking scale is close to the GeV range, indicating the presence of other low-energy states neglected in our effective approach. However, as discussed in great details in Ref.~\cite{Liu:2021wap}, it is actually possible to construct viable UV completions for the ``visible QCD" axion scenario, which is very similar to the effective model, except that the explicit ALP mass term is replaced by the QCD axion mass of similar numerical size. We expect that along the lines of Ref.~\cite{Liu:2021wap} one can devise a viable UV extension of our scenario, although likely rather baroque. Here we refrain from presenting such a model, and focus on the analysis of collider and DM phenomenology in the effective framework, which only has a handful of parameters.  

This paper is organized as follows. In Sec.~\ref{sec:model} we define the basic setup of the model.  The resulting particle phenomenology is analyzed in Sec.~\ref{sec:pheno}, which includes axion decay rates, lepton anomalous magnetic moments,  constraints from beam dump experiments and meson decays and  a discussion of  the necessary flavor alignment to satisfy lepton-flavor violation constraints. Sec.~\ref{sec:darkmatter} is devoted to the DM phenomenology, where we discuss the DM relic abundance, constraints from direct and indirect detection and bounds on DM self-interations. We summarize our conclusions in Sec.~\ref{sec:conclusions}. Further details on direct detection are given in Appendix~\ref{DDdetails}.

\section{Setup}\label{sec:model}

We consider a simplified model with a pseudo-Nambu-Goldstone (pNGB) boson $a$, which only couples to the three SM leptons and a Dirac fermion $\chi$ that will account for DM. The  interaction Lagrangian is given by 
\begin{align}
\label{eq:L}
\mathcal L & =  
 - i  a g_\psi  \overline \psi \gamma_5 \psi    \, , 
\end{align}
with $\psi = e, \mu,\tau, \chi$ and we take all couplings real. This Lagrangian is formally renormalizable, although it is an effective theory since we neglected the radial mode associated with the pNGB. Upon $a$-dependent fermion redefinitions (or using the fermion equations of motion with anomaly terms), we can equivalently describe the relevant interaction terms as
\begin{align}
\label{eq:Lbase2}
\mathcal L & =   \frac{\partial_\mu a}{2 f_a}    c_{\psi} \overline \psi \gamma^\mu\gamma_5 \psi  + c_{\gamma}  \frac{\alpha}{8 \pi }   \frac{a}{f_a}  \epsilon^{\mu \nu \rho \sigma} F_{\mu\nu} F_{\rho \sigma}  \, ,
\end{align}
with $\epsilon^{0123} = -1$. The couplings in Eqs.~\eqref{eq:L} and \eqref{eq:Lbase2} are related by
\begin{align}
C_\psi & \equiv \frac{c_{\psi}}{f_a}   = \frac{g_\psi}{m_\psi}  \, , &    C_\gamma & \equiv \frac{c_{\gamma }}{f_a}  =  \frac{g_e}{m_e} +  \frac{g_\mu}{m_\mu} +  \frac{g_\tau}{m_\tau}\, , 
\label{lagrder}
\end{align}
for $\psi = e, \mu, \tau, \chi$. In total there are 6 parameters, but we will fix $g_\mu$ and $g_\tau$ by reproducing the central value of $(g-2)_\ell$, and $g_\chi$ to reproduce the DM relic abundance, as discussed in the next sections. This leaves as free parameters $m_a$ and $g_e$ controlling the particle phenomenology, while $m_\chi$ is only relevant for DM phenomenology. Besides experimental constraints, this parameter space is subject to the bounds from perturbative unitarity, which put upper bounds on the ALP couplings $g_\psi = m_\psi c_\psi/f_a = m_\psi C_\psi < \sqrt{8\pi/3} \approx 2.9$~\cite{Cornella:2019uxs}.  

Note that in contrast to Refs.~\cite{Alves:2017avw, Liu:2021wap} we do not consider the possibility of a UV contribution to the effective ALP couplings to photons. Such contributions must come from charged fermions chiral under PQ, which can acquire a mass only around the PQ breaking scale, which is much below the electroweak scale in this scenario and thus likely in contrast with experimental constraints. 

In the following we discuss the present particle physics constraints on the parameter space, afterwards we discuss the DM phenomenology.

\section{Particle Phenomenology}\label{sec:pheno}

\subsection{ALP Decays}
We will be interested in ALP masses below ${\cal O} (100)  \MeV$, so that the ALP can only decay into electrons and photons\footnote{As we will discuss in the next section, the ALP cannot decay to DM particles since we need $m_\chi > m_a$, so that the relic abundance  dominantly arises from $p$-wave annihilation $\overline{\chi} \chi \to aa$.}. The corresponding decay rates read
\begin{align}
\Gamma (a \to e^+e^-) & = \frac{m_a } {8 \pi} g_e^2 \sqrt{1 - \frac{4 m_e^2}{m_a^2}} \, , \nonumber \\
\Gamma (a \to \gamma \gamma) & = \frac{\alpha^2 m_a^3} { 64 \pi^3 } \left| C_\gamma^{\rm eff} \right|^2 \, , 
\end{align}
where the effective photon coupling receives contributions from all fermions~\cite{Bauer:2017ris}
\begin{align}
C_\gamma^{\rm eff} & = \sum_{\ell = e, \mu ,\tau} \frac{g_\ell }{ m_\ell} \frac{ 4 m_\ell^2}{m_a^2} f^2 \left(\frac{ 4 m_\ell^2}{m_a^2}\right) \, .
\label{effgammacoupling}
\end{align}
Here, the contribution from SM leptons is defined in terms of the loop function
\begin{align}
  f(x) & = \begin{cases}  \arcsin \frac{1}{\sqrt{x}} & x \ge 1 \\ \frac{\pi}{2} + \frac{i}{2} \ln \frac{1 + \sqrt{1-x} }{1 - \sqrt{1-x} } & x < 1\end{cases} \, , \label{decay}
\end{align}
with the limit 
\begin{align}
x f^2 (x) = \begin{cases}  1 + \frac{1}{3 x} & x \gg 1 \\ \frac{x}{4} \left( \pi + i \ln \frac{4}{x} \right)^2 & x \ll 1 \end{cases} \, .
\end{align}

\subsection{Lepton Anomalous Magnetic Moments}

We determine the ALP couplings to heavy leptons $g_\mu$ and $g_\tau$ in order to reproduce the experimental values for the lepton anomalous magnetic moments  $a_\ell = (g_\ell -2)/2$ for $\ell = e, \mu$, which both deviate from the SM prediction. In the muon sector, the comparison of the (2021) experimental average~\cite{Muong-2:2021ojo} with the SM prediction of the Muon g-2 Theory Initiative~\cite{Aoyama:2020ynm} has pointed to an intriguing 4.2$\sigma$ discrepancy 
\begin{align}
\Delta a_\mu^{\rm disp} (2021) = a_\mu^{\rm EXP} - a_\mu^{\rm SM} =  251 (59) \times 10^{-11} \, .
\end{align}
 However, recent lattice results by the BMW collaboration for the hadronic vacuum polarization (HVP) contribution are in conflict with  dispersive approaches based on low-energy $e^+ e^- \to {\rm hadrons} $ data, and rather suggest the value (using the 2021 experimental result)
\begin{align}
\Delta a_\mu^{\rm lat} (2021) =  107 (70) \times 10^{-11} \, , 
\label{g2muold}
\end{align}
 decreasing the discrepancy to 1.5$\sigma$. Other recent lattice calculations are consistent with the BMW result~\cite{Wang:2022lkq, Ce:2022kxy, ExtendedTwistedMass:2022jpw, FermilabLatticeHPQCD:2023jof, Blum:2023qou}, and also a new measurement of the cross section $e^+ e^- \to \pi^+ \pi^-$ between 0.32 and 1.2 GeV indicates an increase of the HVP contribution in the same direction as lattice data~\cite{CMD-3:2023alj}. 
 
 Given the unclear present situation, here we choose to follow an approach that mediates between the dispersive and the lattice methods, restricting the use of lattice data to the region least prone to systematic uncertainties (the so-called ``window" observable~\cite{Ce:2022kxy}) and using low-energy data otherwise~\cite{Colangelo:2022vok}. With the recent  2023 update of the experimental world average~\cite{Muong-2:2023cdq} this approach gives the value~\cite{Acaroglu:2023cza} 
 \begin{align}
\Delta a_\mu^{\rm wind} (2023) =  181 (47) \times 10^{-11} \, , 
\label{g2mu}
\end{align}
which corresponds to a 3.8$\sigma$ deviation. For a summary of the current status of the $(g-2)_\mu$ SM prediction see Refs.~\cite{Colangelo:2023rqr, muong2}.

In the electron sector instead hadronic contributions are largely irrelevant for the SM prediction, and its uncertainty is mainly driven by the input for the fine-structure constant $\alpha$ (see e.g., Ref.~\cite{Giudice:2012ms} for details). Unfortunately, there are two conflicting experimental determinations of $\alpha$, obtained from spectroscopy of either Cs~\cite{Parker:2018vye} or Rb atoms~\cite{Morel:2020dww}. Using the latest (2022) experimental value for $a_e^{\rm EXP}$ obtained by the Harvard group presented in Ref.~\cite{Fan:2022eto}, we obtain for the discrepancies for the Rb method
 \begin{align}
\Delta a_e^{\rm Rb} =  34 (16) \times 10^{-14} \, , 
\label{g2eRb}
\end{align}
corresponding to a 2.1$\sigma$ deviation. Instead the Cs-method gives a lower SM prediction, resulting in 
 \begin{align}
\Delta a_e^{\rm Cs} =  -102 (26) \times 10^{-14} \, , 
\label{g2eCs}
\end{align}
which corresponds to a 3.9$\sigma$ deviation from the SM prediction. Note that the significance has increased with respect to the older experimental value from 2008, giving deviations of 1.6$\sigma$ (Rb) and 2.4$\sigma$ (Cs). Until the conflicting experimental determinations of $\alpha$ have been clarified, in the following we are simply taking the latter (Rb) result in Eq.~\eqref{g2eRb} at face value. 

\begin{figure*}[t!]
	\centering
	\includegraphics[width=0.7\textwidth]{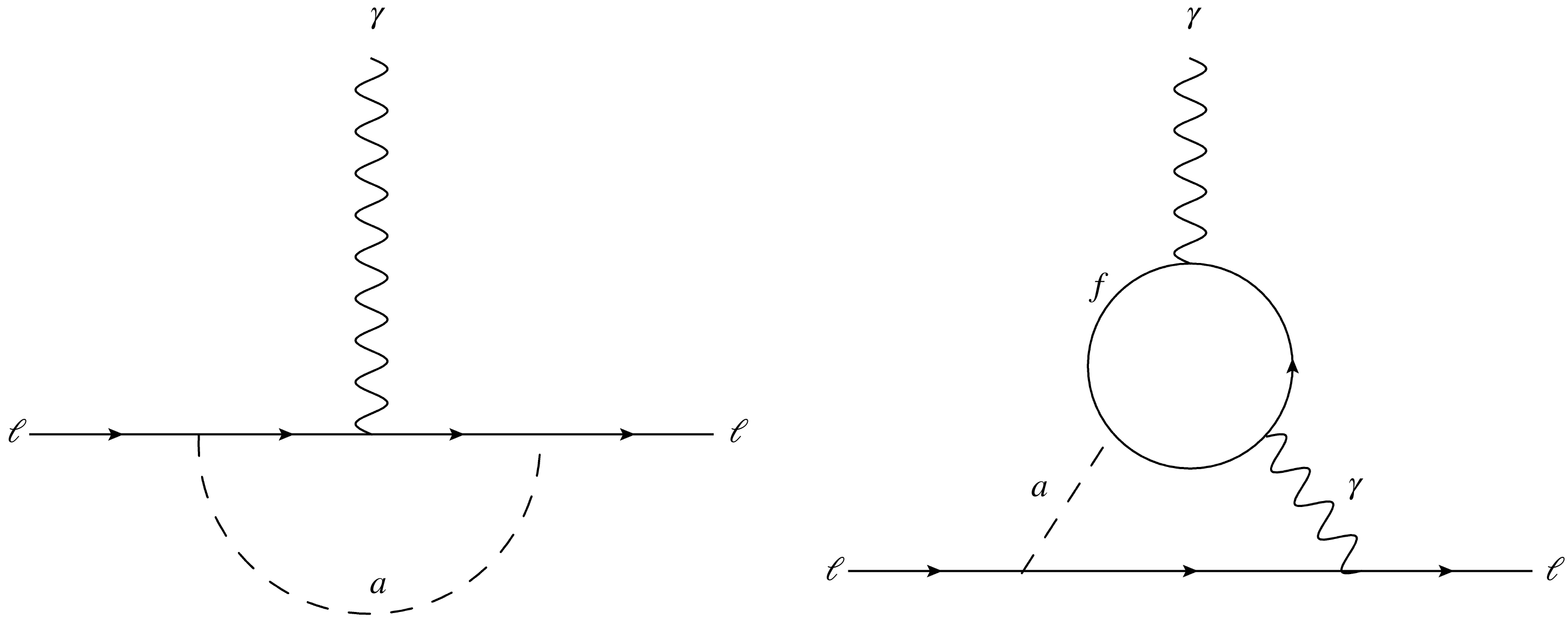}
		  	 \caption{\label{g2diagrams} Feynman diagrams contributing to $(g-2)_\ell$ via the exchange of an ALP coupling to SM leptons.}
\end{figure*}

The Lagrangian  in Eq.~\eqref{eq:L} gives a contribution to the anomalous magnetic moment $\Delta a_{\ell}$ of the lepton $\ell=e,\mu,\tau$ at one-loop~\cite{Cornella:2019uxs, Bauer:2017ris, Chang:2000ii}, corresponding to the diagram on the left in Fig.~\ref{g2diagrams}  
\begin{align}
\Delta a_{\ell}^{\rm 1loop}  & = - \frac{g_{\ell}^2}{16 \pi^2 } h_1\left( \frac{m_a^2}{m_{\ell}^2}\right) \, , 
\label{1loop}
\end{align}
where $h_1(u)$ is a positive-definite loop function given by
\begin{align} 
h_1(u) & = \int_0^1 dy\, \frac{2 y^3}{u- u y + y^2} \, .
\end{align}
Also important are two-loop Barr-Zee diagrams shown on the right in Fig.~\ref{g2diagrams}, which give the following contributions to $\Delta a_{\ell}$~\cite{Buttazzo:2020vfs}
\be
\Delta a_\ell^{\rm 2loop}  =  \frac{\alpha m_\ell }{8 \pi^3 m_f} N_{c}^f Q_f^2 g_{\ell} g_{f} F \left( \frac{m_a^2}{m_\ell^2} , \frac{m_a^2}{m_f^2}\right) \, , 
\label{2loop}
\ee
where $f$ is a fermion with mass $m_f$, color multiplicity $N_{c}^f$, electromagnetic charge $Q_f$ and ALP coupling $g_f$ in Eq.~\eqref{eq:L}, while $F(u, v)$ is the loop function 
\begin{align}
F(u, v) = \int_0^1  dx  \int_0^1 dy \int_0^1 dz \, \frac{u x   }{u  \overline{x} + u v x y z  \overline{z} + v z  \overline{z} x^2  \overline{y}^2 } \,, 
\end{align}
with the shorthand $ \overline{x} = 1-x$, and similar for $y,z$. 

In the limit when the external lepton mass $m_\ell$ is small compared to the ALP mass $m_a$, i.e. $u \gg 1$, we obtain $F(u,v) \to  - \int_0^1 dz   \log (v z\overline{z})/(1- v z \overline{z})$ and thus recover the result in Eq.~(10) of Ref.~\cite{Chang:2000ii} (and Eq.~(58) in Ref.~\cite{Giudice:2012ms}). In the limit of large fermion masses propagating in the loop, $v \ll 1, v \ll u$, one can treat the loop as a pointlike interaction of the ALP with two photons. In this case, our full result should reproduce the leading logarithm obtained from a one-loop calculation within an effective theory, where the heavy fermion has been integrated out. In the limit $v \ll 1 \ll u$ we obtain $F(u, v) \to 2- \log v$, while $v \ll u \ll 1$ gives $F(u, v) \to 3 - \log v/u$, which indeed matches the logarithmic dependence in Eq.~(37) of Ref.~\cite{Bauer:2017ris}, upon identifying the  renormalization scale $\mu$ with the heavy fermion mass $m_f$. 

This discussion also makes clear that in the case  $v \ll 1 \ll u$ the two-loop function  is unsuppressed, in contrast to the one-loop function in Eq.~\eqref{1loop}, which in this limit becomes $h_1(u) \to (-11/3 +2 \log u)/u$. Therefore, the two-loop contribution in Eq.~\eqref{2loop} can potentially dominate over the one-loop contribution in Eq.~\eqref{1loop}, whenever $m_\ell \ll m_a \ll m_f$, even when $g_{f} \sim g_{\ell}$. 

Finally, the total ALP contribution $\Delta a_\ell = \Delta a_\ell^{\rm 1loop} + \Delta a_\ell^{\rm 2loop}$ is compared to the difference of the experimental value and the SM expectation in Eq.~\eqref{g2mu} and \eqref{g2eRb}. We then fix the value of $c_\mu$ and $c_\tau$  such\footnote{There are always two solutions for $C_\mu$, but one of them  involves a large cancellation between one-loop and two-loop contributions. We choose the solution that does not involve such a tuning, so that both central values for $\Delta a_\ell$ are reproduced dominantly by the two-loop ALP contribution. } to reproduce the central values, if possible at all for given values of $m_a$ and $g_e$. The resulting constraints in the $(m_a,  g_e)$ parameter space are shown in Fig.~\ref{phenofig} as excluded regions in light blue.

\subsection{Beam Dump  and Collider Constraints}

\begin{figure}[t!]
	\centering
	\includegraphics[width=\textwidth]{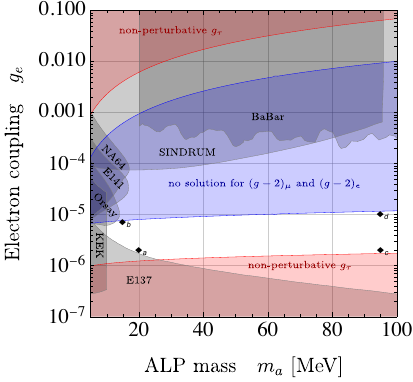}
	  	 \caption{\label{phenofig} Parameter space in the $(m_a,  g_e)$ plane, where ALP couplings to muons and tau leptons are adjusted to reproduce the central values of $\Delta a_\mu$ and $\Delta a_e$. Gray regions show exclusion limits from beam dump and collider experiments,  blue regions denote constraints from anomalous magnetic moments, and red regions are excluded by perturbativity constraints on $g_\tau$. The four benchmark models defined in Table \ref{BMs} are indicated as black diamonds.}
\end{figure}

Stringent constraints on light particles arise from electron beam-dump experiments that have searched for $e^+ e^-$ decays\footnote{We actually assume that the relevant experiments are equally sensitive to ALPs decaying to electrons or photons.} of short-lived particles produced from an electron beam stopped in an absorbing target. Relevant for our scenario are only a handful of experiments. Important bounds on the parameter space are provided by the NA64 collaboration, which originally searched for a massive vector particle~\cite{NA64:2019auh}, and has reanalyzed their results for the case of a pseudoscalar in Ref.~\cite{NA64:2021aiq}. This analysis supersedes previous recasts in Refs.~\cite{Buttazzo:2020vfs,Alves:2017avw} (that were based on a simple coupling rescaling following Ref.~\cite{Bjorken:2009mm}), and gives slightly weaker bounds on ALP couplings to electrons. Also beam dump experiments carried out at SLAC (E141)~\cite{Riordan:1987aw}, KEK~\cite{Konaka:1986cb} and Orsay~\cite{Orsay}  provide relevant constraints on the parameter space, besides limits from the SLAC beam dump E137~\cite{Bjorken:1988as}, which were however shown explicitly only for the lower range of couplings in the original article. For the present scenario instead only the upper range of couplings excluded by E137 is relevant, which we simply take from Fig.~2 of Ref.~\cite{Altmannshofer:2022ckw}. Finally, also colliders have put important bounds on the parameter space. While searches carried out at KLOE~\cite{Anastasi:2015qla} do not provide competitive constraints, the  BaBar collaboration has searched for Dark Photons coupled to electrons, and the resulting limits presented in Ref.~\cite{BaBar:2014zli} can be easily recast for ALPs.  All relevant constraints from beam dumps and colliders are shown as gray regions in Fig.~\ref{phenofig}. 
 
It should be noticed that we have taken into account only experiments that look for ALPs produced off electrons. It is however clear that the ALP also couples to photons at one-loop, and thus constraints from photo-production should also be taken into account. Nevertheless, we expect such constraints to be mild, as the loop contributions to photo-production are suppressed by the relevant energy scale, which is typically large.  This is in contrast to models with effective ALP couplings to photons from loops of heavy fermions, which are severely constrained (see e.g. Ref.~\cite{Bauer:2017ris,Liu:2023bby}). For example, in the relevant mass range of $(10-100)$ MeV strong constraints on effective ALP couplings to photons arise from LEP~\cite{Jaeckel:2015jla}, but we have checked that the resulting constraints on lepton couplings are weaker than constraints from perturbative unitarity. Details on this analysis and constraints on lepton couplings from photo-production will be presented in Ref.~\cite{tobepublished}. 

\subsection{Constraints from Pion Decays}
Important constraints on leptonic ALP couplings also arise from meson decays with ALPs radiated off final-state leptons. The strongest bounds on electron couplings can be obtained from $\pi^+ \to e^+ \nu a, a \to e^+ e^- $ searches at the SINDRUM experiment~\cite{SINDRUM:1986klz}, as recently discussed in Ref.~\cite{Altmannshofer:2022ckw}. This analysis rules out ALP couplings to electrons larger than $10^{-4}$ for  ALP masses below $\sim 50 \MeV$ and ${\rm BR}(a\to  e^+ e^-) \approx 1$, and provides the most relevant upper bound on $g_e$ in the present scenario, see Fig.~\ref{phenofig}, where the parameter region excluded by the SINDRUM search is shown in gray.

\subsection{Flavor Alignment}
A major shortcoming of this scenario is the large amount of required flavor alignment, in order to satisfy stringent constraints from lepton flavor violation. 
Specifically, a possible LFV coupling of the ALP to muons and electrons, ${\cal L} \supset C_{\mu e} \partial_\mu a  \bar \mu \gamma^\mu\gamma_5 e +{\rm h.c.} $, gives rise to LFV muon decays into electrons and ALPs~\cite{Calibbi:2020jvd}, 
\begin{align}
\Gamma (\mu \to e a) & \approx \frac{m_a^3}{16 \pi}  \left| C_{\mu e}\right|^2 \, , 
\end{align} 
where we have neglected corrections of order ${\cal O}(m_e^2/m_\mu^2)$ and ${\cal O}(m_a^2/m_\mu^2)$. Since the ALP promptly decays to $e^+ e^-$, the LFV coupling is subject to the stringent upper limits on ${\rm BR}(\mu \to 3e) \le 3 \cdot 10^{-12}$ by the SINDRUM collaboration~\cite{SINDRUM:1986klz}, which requires
\begin{align}
|C_{\mu e}|  \le \frac{2 \cdot 10^{-12}}{\GeV} \left( \frac{20 \MeV} {m_a}\right)^{3/2} \, , 
\end{align} 
to be compared with $C_e \sim 10^{-2}/\GeV$. For generic nonuniversal PQ charges in the lepton sector one would expect flavor-violating couplings to be of the order of $C_{\mu e} \sim C_e \times \theta^e_{12}$, where $ \theta^e_{12}$ is a mixing angle in the 1-2 sector of charged leptons, parametrizing the misalignment between PQ charges and Yukawa matrices. This mixing angle thus needs to be smaller than roughly $10^{-10}$, which conflicts with simple models of flavor, where the rotation angle is expected to be roughly of the order of $m_e/m_\mu \sim 5 \cdot 10^{-3}$ or  $m_e^2/m_\mu^2 \sim 2 \cdot 10^{-5}$. Thus a nearly perfect alignment of Yukawa and PQ basis is necessary, which might be achievable in models with extended abelian flavor symmetries, analogous to providing flavor alignment in supersymmetric models~\cite{Nir:1993mx}. 
\subsection{Summary}
We summarize the particle physics phenomenology  in Fig.~\ref{phenofig}, which shows the relevant experimental constraints in the $(m_a,  g_e)$ plane, as discussed in this section. It is clear that $(g-2)_\ell$ alone would allow two separated strips of available parameter space, but the upper bound is entirely excluded by SINDRUM searches for $\pi^+ \to e^+ \nu a, a \to e^+ e^- $ in the low mass regime, and by BaBar searches in the high mass range. This leaves a rather narrow strip  with couplings $10^{-6} \lesssim g_e \lesssim 10^{-5}$ and $10 \MeV \lesssim m_a$, smaller masses being excluded by KEK and E137 searches. In the available parameter space we select four benchmark models (BMs), which we take as representatives for the entire region. They are defined in Table \ref{BMs} and will be used to discuss the Dark Matter phenomenology in the next section. 

\begin{table}[h]
\centering
    \begin{tabular}{|c|c|ccc|cc|}
        \hline
        Model & $m_a$[MeV] & $g_e/10^{-5}$ & $g_{\mu}/10^{-4}$ & $g_{\tau}$ & BR$_{\gamma \gamma}$[\%] & $\tau_a$[ps]  \\ \hline
        $a$ & 20 & $0.20 $ & $0.71 $ & $1.8$ & $99 $ & 2.8  \\
        $b$ & 15 &  $0.70$ & $3.6$ & $0.50$ & $20 $ & 18  \\ \hline
        $c$ & 95 & $0.20$ & 0.59 & 2.5 & 100 & 0.01  \\  
        $d$ & 95 & $1.0$ & 3.9 & 0.51 & 84 & 0.28  \\ \hline
    \end{tabular}
    \caption{\label{tabench}  Definition of benchmark models. }
    \label{BMs}
\end{table}
\section{Dark Matter Phenomenology}\label{sec:darkmatter}
We now delve into the connection of  $(g-2)_\ell$ and the DM relic abundance in this scenario. Specifically, among the possible ALP couplings to SM leptons allowed by the constraints of Fig.~\ref{phenofig}, we focus on the four benchmark models as representative examples defined in Table~\ref{BMs}.  The remaining free parameters of the model are  the DM mass $m_\chi$ and the DM-ALP coupling $g_\chi$,  as defined in Eq.~\eqref{eq:L}. We  determine them by requiring that the observed DM abundance is attained after thermal freeze-out of DM fermions from the SM plasma. Subsequently, we assess whether the model is compatible with the current constraints derived from various DM searches.

\subsection{Relic Density}
\begin{figure*}[t]
	\centering
	\includegraphics[width=0.9\textwidth]{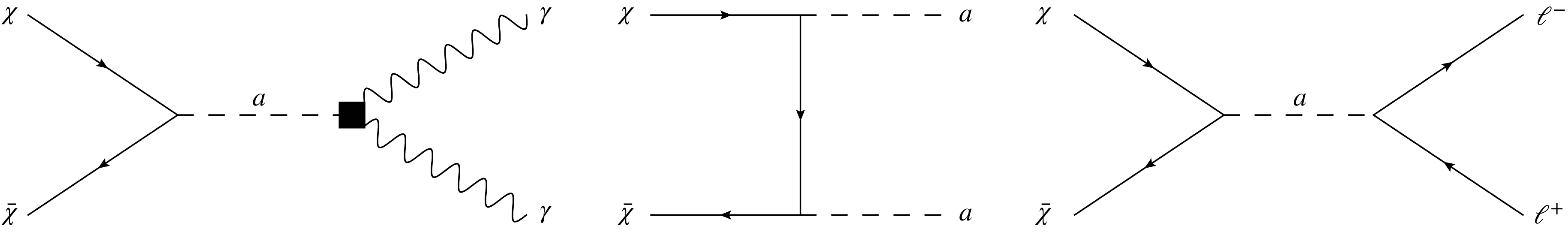}
		  	 \caption{\label{DMdiagrams} Feynman diagrams contributing to DM annihilation. From left to right: $\Bar{\chi} \chi \to \gamma \gamma$, $\Bar{\chi} \chi \to a a$, $\Bar{\chi} \chi \to \ell^+ \ell^-$.}
\end{figure*}

For DM in the mass range $1 \MeV \lesssim m_\chi \lesssim 10 \GeV$, the relevant DM annihilations in the early universe occur through three primary channels: $i)$  tree-level $\Bar{\chi} \chi \to \ell^+ \ell^-$ leptonic channels; $ii)$  tree-level $\Bar{\chi} \chi \to a a$ ALP channel; and $iii)$ the loop-induced $\Bar{\chi} \chi \to \gamma \gamma$ di-photon channel, see Fig.~\ref{DMdiagrams}. The corresponding velocity-averaged cross sections read:
\begin{align}
&\langle \sigma v \rangle_{\ell \ell} = \frac{g_\ell^2 g_\chi^2}{2 \pi} \frac{m_\chi^2}{(4 m_\chi^2-m_a^2)^2} \sqrt{1-\frac{m_\ell^2}{m_\chi^2}} \ , \label{annlep} \\
& \langle \sigma v \rangle_{\gamma \gamma}= \frac{g_\chi^2 \alpha_\text{em}^2}{4 \pi^3} \frac{m_\chi^4}{(4 m_\chi^2-m_a^2)^2} \left| \tilde{C}_\gamma^{\rm eff} \right|^2 \ , \label{annphot} \\
&\langle \sigma v \rangle_{aa}=\frac6x \, \frac{g_\chi^4}{24 \pi}  \frac{m_\chi^2(m_\chi^2 - m_a^2)^2}{(2m_\chi^2-m_a^2)^4} \sqrt{1- \frac{m_a^2}{m_\chi^2}}  \ , \label{annax}  
\end{align}
where $x=m_\chi/T$ and $T$ is the temperature of the thermal bath. The first two cross-sections are $s$-wave, while the third one is $p$-wave. In Eq.~\eqref{annphot}, $\tilde{C}_\gamma^{\rm eff}$ is analogous to the effective photon coupling $C_\gamma^{\rm eff}$ for the ALP decay defined in  Eq.~\eqref{effgammacoupling}, but with the replacement $m_a^2 \to s \simeq 4m_\chi^2$. Following  Ref.~\cite{Servant:2002aq}, we solve the Boltzmann equation for thermal freeze-out including both $s$- and $p$-wave contributions to the total velocity-averaged cross section. Imposing that $\Omega_{\rm DM} h^2\simeq 0.12$~\cite{Planck:2015fie} then determines a line in the ($m_\chi, g_\chi$) plane for a given benchmark model.

\subsection{Indirect Detection}
Indirect detection (ID) searches for DM aim to find hints for  DM by observing anomalies or distinctive features in cosmic ray fluxes measured on earth. These searches typically focus on regions where the DM density is expected to be substantial, such as the centers of galaxies or galaxy clusters. Alternatively, ID searches also target regions with relatively low astrophysical backgrounds, for example dwarf spheroidal galaxies.  Moreover, the Cosmic Microwave Background (CMB) serves as a powerful tool for indirect detection: DM annihilations release energy into the surrounding plasma, causing ionization and heating of the medium~\cite{Padmanabhan:2005es,Cirelli:2009bb}. This injection of energy influences the evolution of the universe during recombination, leaving imprints on the anisotropies of the CMB and resulting in distortions of the CMB blackbody spectrum~\cite{Chluba:2011hw,Slatyer:2015jla}.

We focus on two robust indirect detection probes: $(i)$ CMB anisotropies, which are highly effective in constraining sub-GeV DM annihilations into electrons and di-photons, relevant for our scenario. For a given annihilation channel we take these bounds from Ref.~\cite{Slatyer:2015jla}; and $(ii)$ a compilation of X-rays and $\gamma$-ray data from several experiments (see e.g., Refs.~\cite{Essig:2013goa,Cirelli:2023tnx}). In particular, in the DM mass range spanning from hundreds of MeV to tens of GeV, the bounds coming from DM searches towards dwarf spheroidal galaxies~\cite{Fermi-LAT:2015att,Hess:2021cdp} surpass the CMB constraints only if the annihilation channel into heavy leptons ($\mu, \tau$) is accessible. 
The constraints derived from the CMB, when compared to those obtained from dwarf galaxies~\cite{Fermi-LAT:2015att,Hess:2021cdp}, are generally considered more robust. This is due in part to the complexities involved in computing $J$-factors, as extensively discussed in~\cite{Lefranc:2016dgx}, which contribute to an overall systematic uncertainty of 0.76 dex. As a result, the constraints on the coupling parameter $g_{\chi}$ may be affected, potentially weakening by up to a factor of 2.5, particularly for $m_{\chi} \gtrsim m_{\tau}$. Nevertheless, these uncertainties do not significantly alter the main findings of our study depicted in Fig.~\ref{DMfig}.
Furthermore, it is important to note that above 180 MeV the constraints from the XMM-Newton satellite, provided in Ref.~\cite{Cirelli:2023tnx}, are the strongest available. Nevertheless, these constraints are subject to considerable uncertainties, as shown in Fig.~9 of~\cite{Cirelli:2023tnx}, and therefore, we opt not to include them in our analysis.

In the aforementioned probes the DM is highly non-relativistic. Consequently, the $p$-wave DM annihilations into ALPs are  always negligible when compared to the $s$-wave annihilations into pairs of leptons and di-photons.    Hence, to compute the limits from indirect detection we collect the relevant bounds and weigh them by the  ratios of the respective $s$-wave channels. Specifically, we first calculate the total velocity averaged  $s$-wave cross section $\langle \sigma v \rangle^\text{th}_\text{tot}=\sum_{i} \langle \sigma v \rangle_{ii} $ and the ratios $\mathcal{R}_{i} =  \langle \sigma v \rangle_{ii} / \langle \sigma v \rangle^\text{th}_\text{tot}$ with $i=\gamma,e,\mu,\tau$. Then by denoting $ \langle \sigma v \rangle^\text{CMB}_{ii}$ and $\langle \sigma v \rangle^\text{Fermi}_{\tau^+\tau^-}$ as the best experimental limits in a given channel, we infer a limit in the $(m_\chi, g_\chi)$ plane  by demanding that
\begin{equation}
\langle \sigma v \rangle^\text{th}_\text{tot} < \sum_{i=\gamma,e,\mu}\mathcal{R}_{i}   \langle \sigma v \rangle^\text{CMB}_{ii}  +\mathcal{R}_{\tau}   \langle \sigma v \rangle^\text{Fermi}_{\tau\tau} \, .
\end{equation}
\subsection{Direct Detection}
One of the notable advantages of the present scenario is that constraints from direct detection (DD) searches for DM are easily satisfied. Model-independent processes originating from either single or double ALP exchange between the DM fermion and SM leptons give rise to loop-induced scatterings of DM off nuclei, which are highly suppressed and therefore remain entirely undetectable within current and upcoming DD experiments.

The only plausible process that could result in detectable collisions with SM quarks is induced by the model-dependent trilinear vertex $A_{aah} a^2 h$, which couples the ALP to the SM Higgs. While not included in our setup defined in Eq.~\eqref{eq:L}, such a coupling is likely generated within UV-complete models.  As discussed in the introduction, a possible UV completion of our effective framework is provided in Ref.~\cite{Liu:2021wap}, where an ALP (actually a proper QCD axion) with mass in the MeV range and sizable couplings to SM fermions emerges. In this particular setup, the dimensionful coupling $A_{aah}$ is of the order of hundreds of MeV. According to the results in Ref.~\cite{Ipek:2014gua}, the vertex $A_{aah} a^2 h$ then leads to a sizable spin-independent DM-nucleon cross section, which is already excluded by the latest results from the LUX-ZEPLIN (LZ) experiment~\cite{LZ:2022lsv}, if the DM mass exceeds roughly tens of GeV (see Appendix~\ref{DDdetails} for more details). For this reason we restrict the parameter space to $m_\chi \le 10$ GeV, although it might be possible to construct UV completions, where the trilinear coupling $A_{aah}$ is much smaller.
\subsection{Other Complementary Bounds}
In addition to  these purely observational constraints, we also consider two other complementary bounds on the ALP-DM coupling from: $(i)$ perturbative unitarity which gives $g_\chi < \sqrt{8 \pi/3}$~\cite{Cornella:2019uxs}; and $(ii)$ DM self-interactions which gives $g_{\chi} \lesssim 0.21 \left(m_\chi/\MeV \right)^{3/4}$~\cite{Dror:2023fyd}. This limit arises from the fact that collision processes like $\chi \chi \to \chi \chi$, $\Bar{\chi} \Bar{\chi} \to \Bar{\chi} \Bar{\chi}$ and $\chi \Bar{\chi} \to \chi \Bar{\chi}$ can in principle transport heat from the hotter outer region to the colder inner region of the DM halo, leading to a thermalization of the latter if the energy transfer cross section per unit mass $\sigma_T/m_\chi$ is larger than roughly $1 \ \text{cm}^2 / \text{g}$~\cite{Spergel:1999mh,Tulin:2013teo}.

\begin{figure*}[t!]
	\centering
	\includegraphics[width=.45\textwidth]{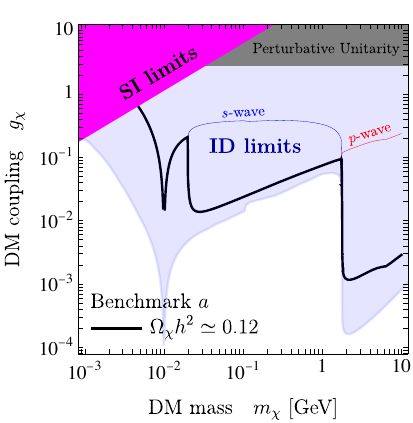} \
	 \includegraphics[width=.45\textwidth]{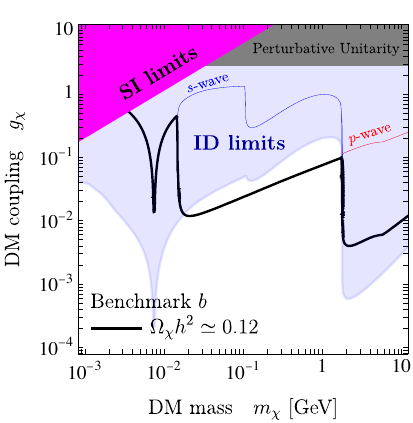} \\
	  \includegraphics[width=.45\textwidth]{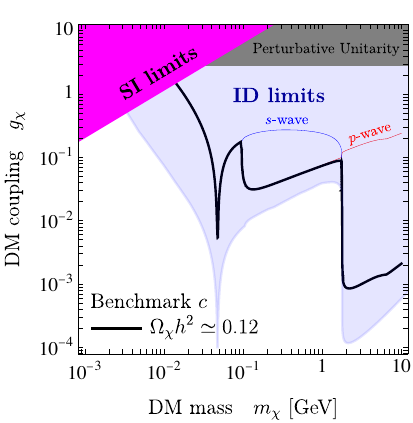} \
	   \includegraphics[width=.45\textwidth]{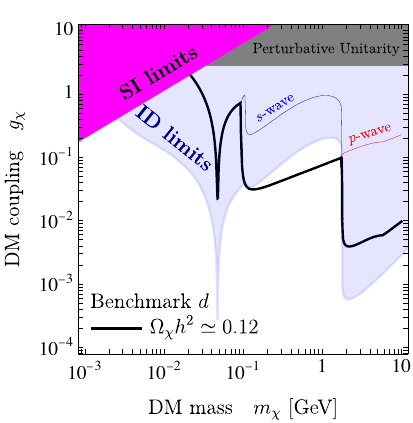}
	  	 \caption{\label{DMfig} Parameter space in the $(m_\chi, g_\chi)$ plane. The left column displays  benchmark  models $a$ and $c$, while the right column shows benchmarks $b$ and $d$. In all panels, the thick black line corresponds to the portion of the parameter space where the observed DM relic abundance is thermally produced (the $s$- and $p$-wave contributions to the relic density are depicted as thin red and blue lines, respectively). The constraints from various DM searches are represented as colored shaded regions.  }
\end{figure*}

\subsection{Summary}
In Fig.~\ref{DMfig}, we present a summary of the results in the ($m_\chi, g_\chi$) plane. The left column displays  benchmark models $a$ and $c$, while the right column shows benchmarks $b$ and $d$. In all panels, the $s$- and $p$-wave contributions to the relic density are depicted as thin red and blue lines, respectively. The  thick black line corresponds to the portion of the parameter space where the observed DM relic abundance is thermally produced. The blue shaded regions represent the bounds derived from DM indirect detection. Additionally, we use dark gray shading to indicate areas of parameter space that are excluded by the requirement of perturbative unitarity. Regions excluded by limits obtained from DM self-interactions are highlighted with dark magenta shading. 

\medskip
In the mass range $m_e<m_\chi<m_a$, both $p$-wave annihilations into ALPs and  $s$-wave annihilations into heavy leptons are kinematically closed. Since in all benchmarks $C_\tau$ is the largest coupling among the leptonic couplings $C_\ell$, it dominates the effective photon coupling in Eq.~\eqref{annphot}, $|\tilde{C}^{\rm eff}_\gamma| \simeq  C_\tau = g_\tau/m_\tau$. Hence,  the ratio of the thermally averaged cross sections in Eq.~\eqref{annlep} and Eq.~\eqref{annphot} is given by  $\langle \sigma v \rangle_{ee}/\langle \sigma v \rangle_{\gamma\gamma} \simeq 2\pi^2/\alpha_{\rm em}^2 \, g_e^2/g_\tau^2 \, m_\tau^2/m_\chi^2$. Using the reference values for the four benchmark models in Table~\ref{tabench}, it is  evident that for BMs $a$ and $c$, the dominant contribution to the relic density arises from loop-induced annihilations into photon pairs. Conversely, for  benchmarks $b$ and $d$, tree-level annihilations into pairs of electrons are the primary channel (if the axion is not too heavy, $m_a \lesssim 60 \MeV$, otherwise again annihilation into photons dominates). In this specific mass range and for all four benchmark models, the CMB limits probe $s$-wave cross section into either di-photons or electron pairs that are considerably smaller than the thermal cross-section. Thus, for $m_\chi<m_a$, the CMB limits completely rule out the freeze-out predictions depicted in Fig.~\ref{DMfig}.

\smallskip
In the mass range $m_a<m_\chi<m_\tau$, the relic abundance is dominantly generated by  $p$-wave annihilation into ALPs. This holds for all models and is particularly important, because for sub-GeV thermal DM with $p$-wave annihilations indirect detection constraints are less restrictive.  Specifically, for BMs $a$ and $c$ the ratio between the $p$-wave cross section at freeze-out $\left.\langle \sigma v \rangle_{aa}\right|_{x=2x_{\rm F}}\simeq 1/x_{\rm F}\,g_\chi^4/(128\pi m_\chi^2)$ with $x_{\rm F}\simeq 25$ and the $s$-wave cross section into di-photons is $\left.\langle \sigma v \rangle_{aa}\right|_{x=2x_{\rm F}}/\langle \sigma v \rangle_{\gamma\gamma}\simeq  \pi^2/(2x_{\rm F} \alpha_{\rm em}^2) \, g_\chi^2/g_\tau^2 \, m_\tau^2/m_\chi^2$. For the reference values in Table~\ref{tabench}, it is clear that the cross section into diphotons is approximately one order of magnitude smaller than the $p$-wave cross-section at freeze-out for DM for $m_\chi \simeq 1$ GeV. It is worth noticing that in this specific mass range, $\langle \sigma v \rangle_{\gamma\gamma}$ remains independent of $m_\chi$.  For DM masses in the GeV range the CMB limits are particularly stringent. They rule out cross sections that are around 30 times smaller than the thermal value, and this exclusion becomes more pronounced as the DM mass decreases. Consequently, the freeze-out predictions  for BMs $a$ and $c$, depicted by the solid black lines in the left column of Fig.~\ref{DMfig}, are robustly excluded. For BMs $b$ and $d$, when $m_\chi < m_\mu$ the primary $s$-wave channel is still annihilation into electron pairs, while  for $m_\chi > m_\mu$ the muon channel becomes increasingly relevant. This transition results in an enhanced $s$-wave contribution to the relic density, as depicted by the blue lines in the right column of Fig.~\ref{DMfig}. In this specific mass range  $\left.\langle \sigma v \rangle_{aa}\right|_{x=2x_{\rm F}}/\langle \sigma v \rangle_{\ell\ell}\simeq 1/(4x_{\rm F}) \, g_\chi^2/g_\ell^2 $. Hence, for the reference values in Table~\ref{tabench}, the $s$-wave channel into electrons is significantly smaller than the thermal value, at least by a factor of $10^4$. Conversely, the $s$-wave channel into muons is also notably smaller, but by a factor of at least 50 when compared to the thermal value. The indirect detection limits for leptonic channels are not yet sensitive enough to probe cross sections of such small size. Consequently, the thermal freeze-out predictions for the benchmark models in the right column of Fig.~\ref{DMfig} remain viable and are not excluded by these limits.

\smallskip
Finally, for $m_\chi>m_\tau$, the $s$-wave annihilation into tau leptons is open. Since $g_\tau$ in all benchmarks  is close to unity, this process emerges as the dominant mechanism for generating the DM abundance in the multi-GeV  mass range and beyond. In this specific mass range the indirect detection limits from DM searches that are directed towards dwarf spheroidal galaxies are extraordinarily stringent. These limits are so severe that they completely rule out the possibility of thermal DM with $s$-wave annihilations into tau lepton pairs up to DM masses of hundreds of GeV.

\section{Summary and Conclusions}\label{sec:conclusions}
We have discussed a simple scenario to connect the observed discrepancies in anomalous magnetic moments with the Dark Matter relic abundance. In this framework an axion-like particle in the multi-MeV range couples solely to SM leptons and a DM fermion. In the lepton sector, the couplings to muons and tau leptons are fixed by explaining the discrepancies in $(g-2)_\ell$, leaving as free parameters only the ALP mass and its electron coupling. Imposing the present constraints from beam-dumps and collider experiments, the viable parameter space consists of a narrow strip with $10^{-6} \lesssim g_e \lesssim 10^{-5}$ (and $10 \MeV \lesssim m_a $), see Fig.~\ref{phenofig}. We have identified four benchmark models that represent this parameter space, and contribute to $(g-2)_\mu$ and  $(g-2)_e$ dominantly through 2-loop Barr-Zee diagrams. 

The remaining parameter space in the dark sector, i.e. the mass of the DM fermion and its coupling to the ALP is determined by requiring that the observed DM relic abundance is reproduced through thermal freeze-out. This determines a line in this 2D parameter region for a given benchmark model, which is subject to constraints from direct and indirect detection, see Fig.~\ref{DMfig}.  We find that CMB constraints completely rule out $s$-wave annihilation channels into electrons, muons or photons, while $s$-wave annihilation into tau leptons is excluded by direct detection constraints. This leaves as the only viable possibility $p$-wave annihilation into ALPs, which essentially fixes the DM mass range $m_a < m_\chi < m_\tau$, along with the DM-ALP coupling of order few $\times 10^{-2}$. Still, even in this region ALP couplings to photons have to be sufficiently suppressed in order to satisfy CMB constraints, which disfavors large ALP-tau couplings, and in turn small electron couplings, so only the upper part of the phenomenologically allowed strip in the parameter space is compatible with DM phenomenology. This leaves only a relatively narrow region that will be tested by current and future experiments, such as XMM-Newton searching for X-rays generated via inverse Compton processes, the next generation of CMB probes and possibly dedicated searches at beam dump experiments.
\section*{Acknowledgments}
The research conducted by G.A. and P.P. receives partial funding from the European Union–Next generation EU (through Progetti di Ricerca di Interesse Nazionale (PRIN) Grant No. 202289JEW4). J.W. is part of the International Max Planck Research School (IMPRS) on ``Elementary Particle Physics". We thank Giorgio Arcadi, Jeff Dror, Francesco d'Eramo, Diego Redigolo and Stefan Vogl for useful discussions.

\appendix
\section{Details of Direct Detection Constraints}\label{DDdetails}
\begin{figure}[h!]
	\centering
	 \includegraphics[width=\textwidth]{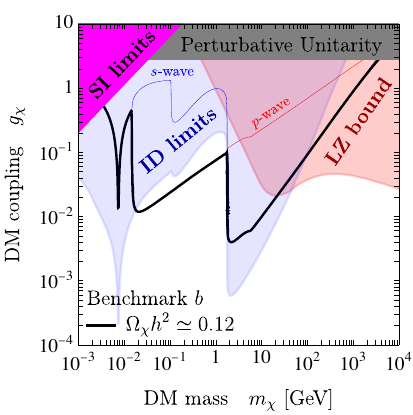} \
	  	 \caption{\label{DMfigzoomed} Enlarged parameter space for benchmark model $b$.  In addition to the previously discussed bounds, we also include the model-dependent direct detection constraint from LZ as a shaded light red region.}
\end{figure}

In terms of the model parameters of Ref.~\cite{Ipek:2014gua}, the trilinear vertex $a^2h$ has an effective coupling $A_{aah}=2 m_A^2 \theta^2/v$ in the limit $m_A\gg m_a$. A rough estimate of the DM-nucleon spin-independent cross section induced by this interaction, expressed in terms of $A_{aah}$, is
\begin{equation}\label{sigmaSI}
\begin{split}
\sigma_{\rm SI}   \simeq & \,  5.2 \cdot 10^{-43} \text{ cm}^2 \times \\ 
& \left(\frac{A_{aah}}{\text{1\,GeV}}\right)^2 \left(\frac{\text{100\,MeV}}{q_{\rm ref}}\right)^4 \left(\frac{g_\chi}{0.5}\right)^4  \left(\frac{m_\chi}{\text{30\,GeV}}\right)^2  \ ,
\end{split}
\end{equation}
 where $q_{\rm ref} = 2 \mu_{\chi,\rm Xe} \, v \simeq (1 \div 100) \MeV \approx m_a$ is the typical momentum exchanged in DM collisions off xenon nuclei and we have used the $\bar q  q$ matrix element $\langle N | \sum_q m_q \bar q q | N \rangle=105.9$ MeV from  the FLAG average of the lattice computations in the case of $N_f = 2 + 1 + 1$~\cite{FlavourLatticeAveragingGroup:2019iem}. 
An estimate of the trilinear coupling can be obtained from Eqs.~(43-44) of Ref.~\cite{Liu:2021wap} which yields  $A_{aah}\simeq 0.28$ GeV. With this input we can calculate the limits in the ($m_\chi,g_\chi$) plane by comparing Eq.~\eqref{sigmaSI} with the LZ bound of the DM spin-independent cross section~\cite{LZ:2022lsv}. This limit is roughly the same for all benchmark models and we choose to illustrate it specifically for model $b$ in Fig.~\ref{DMfigzoomed}. As it is apparent, the LZ limit rules out the possibility of thermal DM for $m_\chi$ larger than tens of GeV.

\bibliographystyle{utphys}
\bibliography{Bib}

\end{document}